\documentclass[traditabstract]{aa}
\usepackage{subfigure}
\usepackage{longtable}
\usepackage{graphicx}
\usepackage{graphics}
\usepackage{keyval}
\usepackage{trig}
\usepackage{psfig,epsf}
\usepackage{txfonts}
\def\beq{\begin{equation}}
\def\eeq{\end{equation}}

\begin{document}

\title{Disk mass and disk heating in the spiral galaxy NGC~3223\thanks{Based on observations
    collected at the European Southern Observatory, Chile, under
    proposal 68.B-0588.}}

\author{G. Gentile\inst{1,2} \and C. Tydtgat\inst{1,3} \and
  M. Baes\inst{1} \and G. De Geyter\inst{1} \and M. Koleva\inst{1} \and G. W. Angus\inst{2} \and W. J. G. de
  Blok\inst{4,5,6} \and W. Saftly\inst{1} \and S. Viaene\inst{1}}  

\institute{Sterrenkundig Observatorium, Universiteit Gent, Krijgslaan 281, B-9000 Gent, Belgium\\
              \email{gianfranco.gentile@ugent.be}
    \and
            Department of Physics and Astrophysics, Vrije Universiteit
            Brussel, Pleinlaan 2, 1050 Brussels, Belgium
   \and
            Department of Solid State Sciences, Krijgslaan 281, B-9000 Ghent, Belgium
  \and
          Netherlands Institute for Radio Astronomy (ASTRON), Postbus 
          2, 7990 AA Dwingeloo, the Netherlands
   \and 
          Astrophysics,
 Cosmology and Gravity Centre, Department of Astronomy, University of
 Cape Town, Private Bag X3, Rondebosch 7701, South Africa
   \and
    Kapteyn Astronomical Institute, University of
 Groningen, PO Box 800, 9700 AV Groningen, the Netherlands
      }

\abstract
{
We present the stellar and gaseous kinematics of an Sb galaxy,
NGC~3223, with the aim of determining the vertical and radial stellar
velocity dispersion as a function of radius, which can help to
constrain disk heating theories. Together with the
observed NIR photometry, the vertical velocity dispersion is also used
to determine the stellar mass-to-light (M/L) ratio, typically one of the largest
uncertainties when deriving the dark matter distribution from the
observed rotation curve. We find a
vertical-to-radial velocity dispersion ratio of
$\sigma_{\rm z}/\sigma_{\rm R}=1.21 \pm 0.14$, significantly higher 
than expectations from known correlations, and a weakly-constrained Ks-band stellar M/L ratio in
the range 0.5--1.7, at the high end of (but consistent with) the
predictions of stellar population synthesis models. Such a weak
constraint on the stellar M/L ratio, however, does not allow us to
securely determine the dark matter density distribution. To achieve
this, either a statistical approach or additional data
(e.g. integral-field unit) are needed.
}


\maketitle

\section{Introduction}
\label{sec_intro}

The structure of galaxies, and in particular their mass content and
distribution, has been extensively investigated in the past decades.
Studying the mass distribution of galaxies is closely related to the
disk mass problem: how does one associate the observed stellar light
distribution to a certain stellar mass distribution? Typically this is
achieved by scaling the stellar light distribution by a factor (the
mass-to-light ratio, M/L) to derive the stellar mass
distribution. However, the stellar M/L ratio is not known a priori,
and a number of routes exist to estimate it
(De Jong \& Bell 2007, Courteau et al. 2014): for instance, via stellar population
synthesis models (Bell et al. 2001, McGaugh \& Schombert 2014), spiral arms (Fuchs 2003), bar
dynamics (Weiner et al. 2001, P{\'e}rez et al. 2004), and, most
importantly for the present paper, using vertical velocity dispersions
(Bottema 1997, Kregel et al. 2005, Bershady et al. 2010a, Martinsson et
al. 2013). These methods give only very roughly consistent results,
discrepancies can run up to factors of a few.
A closely related issue is the maximum disk hypothesis: can the stellar
disk contribution to the rotation curve be scaled up to its maximum
possible value (van Albada et al. 1985)? Using the various methods
described above to derive the stellar mass, authors have reached
contrasting conclusions, ranging from close to maximum disk to
submaximal disks (e.g. Weiner et al. 2001, Martinsson et al. 2013). 

The determination of the stellar mass is important for the derivation
of the dark matter distribution in galaxies, notably because the shape
of the dark matter density distribution is related to that of the stellar
mass (de Blok et al. 2008). The dark matter distribution has been the
topic of a debate in the last years, where it has become
apparent that there is a discrepancy between the dark matter density
profiles found in dark matter-only $\Lambda$ Cold Dark Matter
($\Lambda$CDM) simulations of galaxy formation and the observed dark
matter density profiles (e.g. de Blok et al. 2001, Gentile et al. 2004, van
Eymeren et al. 2009, Kuzio de Naray \& Kaufmann 2011). Recently,
various models involving the (unknown) effect of baryons on the dark
matter distribution have been proposed, trying to reduce or remove the
discrepancy between simulations and observations (e.g. Oh et al. 2011a,
Governato et
al. 2012, Macci{\`o} et
al. 2012, Cloet-Osselaer et
al. 2012, Di Cintio et al. 2014).

The internal kinematics of galaxies have been used 
to investigate a wide variety of topics, including the mass
distribution of galaxies (e.g. Bosma 1981, Persic et al. 1996, Oh et
al. 2011b, Gentile et al. 2013) and galaxy evolution, in particular
disk heating (e.g. Minchev \&
Quillen 2006, Gerssen \&
Shapiro Griffin 2012, Sellwood 2014).
A possible route to estimate the stellar mass is to
use the relation between vertical stellar velocity dispersion and
surface density, as done by the DiskMass survey team (Bershady et
al. 2010a). Stellar kinematics can also be used to investigate the process called
``disk heating'', the increase of the random velocity of stars in the
galaxy disk with time. By measuring the vertical and
radial velocity dispersions, $\sigma_{\rm z}$ and $\sigma_{\rm R}$, one can test the
validity of theories put forward to explain the disk heating
phenomenon (Jenkins \& Binney 1990, Shapiro et al. 2003), e.g. spiral
structure (Sellwood \& Binney 2002), scattering off giant molecular
clouds (H\"anninen \& Flynn 2002), or the infall of substructure onto
the galaxy (e.g. Kazantzidis et al. 2009). Note that there is still no consensus
concerning the dominant mechanism(s) at play.
Gerssen \&
Shapiro Griffin (2012) find that the variation of $\sigma_z$ and
$\sigma_{\rm R}$ with Hubble type suggest that spiral structure might be
the primary radial heating agent, while giant molecular clouds are
consistent with being the three-dimensional heating agent. These
results are based on a handful of galaxies, and the values of of
$\sigma_{\rm z}$ and $\sigma_{\rm R}$ are derived through a multi-parameter fit of
the observed velocity dispersion and rotation velocity profiles. In
the present paper we use a similar approach.

To tackle simultaneously the issues of disk mass and disk heating,
as a pilot study
we observed the Sb spiral galaxy NGC~3223, whose Hubble distance
(corrected for Virgocentric flow) is about 38 Mpc. NGC~3223 was
selected because of its southern declination (which made it accessible
for observations with the VLT), its moderate inclination and
relatively smooth optical appearance. It was part of the sample of 74 galaxies
for which Palunas \& Williams (2000) made 2-D H$\alpha$ velocity
fields, and then derived a rotation curve. 
We obtained H{\sc i} data, to
extend the rotation curve to larger radii, and stellar kinematics along
the minor and major axes.
NGC~3223 was also part of the sample of Kassin et al. (2006a,b) who
used the rotation curve of Mathewson et al. (1992) to derive the
baryonic and dark matter distributions in a sample of galaxies.

The paper is structured as follows: in Section \ref{sec_data} we
present the data we used and the reduction process, then in Section
\ref{sec_analysis} we describe the way we analysed the data. The main
results are shown in Section \ref{sec_results} and then in Section
\ref{sec_conclusions} we draw our conclusions.

\section{Data acquisition and reduction}
\label{sec_data}

\subsection{Photometry data}

The BVI frames were taken at La Silla using the NTT telescope (New
Technology Telescope), mounted with the SUSI-2 detector (D'Odorico et
al. 1998),
during the night of 3$^{\rm rd}$ and 4$^{\rm th}$ of February
2002. Ten frames were taken in each 
band, with an exposure time of 60 seconds (B-band) and 30 seconds (V-
and I-band). The frames were cleaned of cosmic ray artifacts,
de-biased, flat-fielded and sky subtracted with the {\sc MIDAS}
software package. The images were then
aligned and combined together. For flux calibration, the extinction
coefficients and zero points were determined using Landolt standard
star fields SA98 and SA104(b) . The images of these star fields were
reduced like the BVI photometric images. Sky subtraction is
automatically performed when the magnitude is determined in {\sc
  MIDAS}. A flux calibration check was done
by comparing measured fluxes with data found using the Aladin sofware
package (Bonnarel et al. 2000).  Finally,
an astrometric calibration was done using the Gaia package from
Starlink.   

The H- and Ks-band images were made with the infrared imager-spectrometer
SOFI (Son OF Isaac, Moorwood et al. 1998). Ten frames were taken through the
H-filter and ten frames through the Ks-filter, with an exposure time
of 10 seconds. Between each frame an
offset frame was taken to determine the sky background. The infrared sky background
varies rapidly so, following the SOFI manual, the sky background
images were made using a jitter template from which a median
background is determined and subtracted from the NGC~3223
frames. Flat-fielding and illumination correction was done using the
illumination and correction fields for the H- and Ks-band downloaded
from the ESO SOFI website page. 
Flux calibration was performed using the standard starfields SJ9111
and SJ9144 from which the zero point and extinction coefficients were
determined using a growth curve and SNR analysis. The same way as for
the BVI frames, an
astrometric calibration was performed with the Gaia package from Starlink. 
 
\subsection{Spectroscopic data}

The spectroscopic observations were executed during the 6$^{\rm th}$and 7$^{\rm th}$ of
January 2012, for a total exposure time of
6.5 hours. 
Long slit spectra were taken along the major (PA 51 deg) and the minor axis ($-$39\,deg) of
NGC~3223 with the FORS (Focal Reducer and low dispersion
Spectrograph, Appenzeller et al. 1998) detector mounted on the VLT telescope using a 0.7\,arcsec 
wide slit in combination with the Gris\_1028z grism. This resulted in 
an instrumental velocity dispersion of $\sim$ 20\,km s$^{-1}$. 
The covered wavelength range is 7663-9335\,\AA. while the seeing fluctuated around 0.85\,arcsec.
Standard data reduction was
performed with MIDAS. It included bias subtraction,
flat-fielding, cosmic ray removal and wavelength calibration. 
Finally, the spectra were aligned and averaged and sky subtraction was performed.
The 2-D error spectra were computed by taking into account the gain and the read out 
noise of the detector.  

For the data analyses we used
ULySS\footnote{\url{htttp://ulyss.univ-lyon1.fr}}
(Koleva et al. 2009). UlySS is a full spectrum fitting package
that fits the observed spectrum against a linear combination of
single stellar population models, parametrised by their ages and metallicities. 
It takes into account the
extinction and the imperfections of the flux calibration (or lack
thereof) by
including multiplicative polynomial to the fit. In this case we used a
6$^{\rm th}$ degree polynomial.  As base models we use Vazdekis single
stellar population models (SSPs) (Vazdekis et al. 2010) coupled with the CaT library
from Cenarro et al. (2001). Those were convolved with the line-of-sight velocity
distribution function (LOSVD). Thus, the free parameters in the fit
were the age and metallicity of the SSPs, the velocity and the velocity
dispersion of the LOSVD. We used the wavelength range between 8300\AA\
and 8680\AA, which is relatively clear of sky lines.

To access the instrumental broadening and the precision of the
wavelength calibration we used ULySS to fit Gaussians to the reduced
lamp calibration image. The instrumental dispersion we derived changes
from 25\,km s$^{-1}$  at 7900\AA~to 15\,km s$^{-1}$  at 9200\AA. This instrumental
dispersion roughly corresponds to the 22.5\,km s$^{-1}$  broadening of the SSP
models, so we could directly use the output of the program as a measure
of the galaxy's physical velocity dispersion. The precision of the
wavelength calibration was estimated to be around 1\,km s$^{-1}$ . 
For the radial profile analyses we binned the spectra to a
minimum mean signal-to-noise of 15, except for the outer points where
we lowered our requirements to a signal-to-noise ratio of 5. 

\subsection{H{\sc i} data}

The H{\sc i} observations were performed at the ATCA (Australia
Telescope Compact Array) with three different array setups: the 6km
configuration on May 11, 2002, the 1.5km configuration on July 21-22,
2002, and the 750m configuration on August 5-6, 2002, for a total of
$\sim 1700$ minutes on source. The correlator setup was chosen to have
512 16 kHz channels with a total 
bandwidth of 8 MHz.

The data were reduced using the Miriad software package (Sault et
al. 1995). We used standard procedures to flag, calibrate, and
subtract the continuum. The uv-data were then Fourier inverted (using
a Gaussian taper with a width in the image plane of 10 arcsec and a
``robust'' parameter of 0.5), and we
obtained a data cube. The ``dirty'' cube was Hanning-smoothed, yielding a
velocity resolution of 6.6 km s$^{-1}$ and an rms noise of 1.2 mJy
beam$^{-1}$ (corresponding to a column density of 1 $\times$ 10$^{19}$
atoms cm$^{-2}$). We then CLEANed  the data cube and restored it with a
Gaussian beam of FWHM 36.94 $\times$ 23.59 arcsec. 

\section{Data analysis}
\label{sec_analysis}

\subsection{Photometry}
\label{sec_photometry}

The BVIHKs images were further analysed using the software packages FitSKIRT 
(which uses genetic algorithms to fit optical and NIR images, including
radiative transfer calculations, De Geyter et al. 2013), and {\sc BUDDA}
(BUlge Disc Decomposition Analysis, de Souza et al. 2004, which was
designed to perform 2D bulge-disk decompositions of optical and NIR
images). We performed
a 2D bulge-disk decomposition of the Ks-band image because it traces best the
stellar disk mass. We modelled our image as a S\'ersic bulge plus an exponential
disk.

\begin{figure}
\begin{center}
\includegraphics[width=0.47\textwidth]{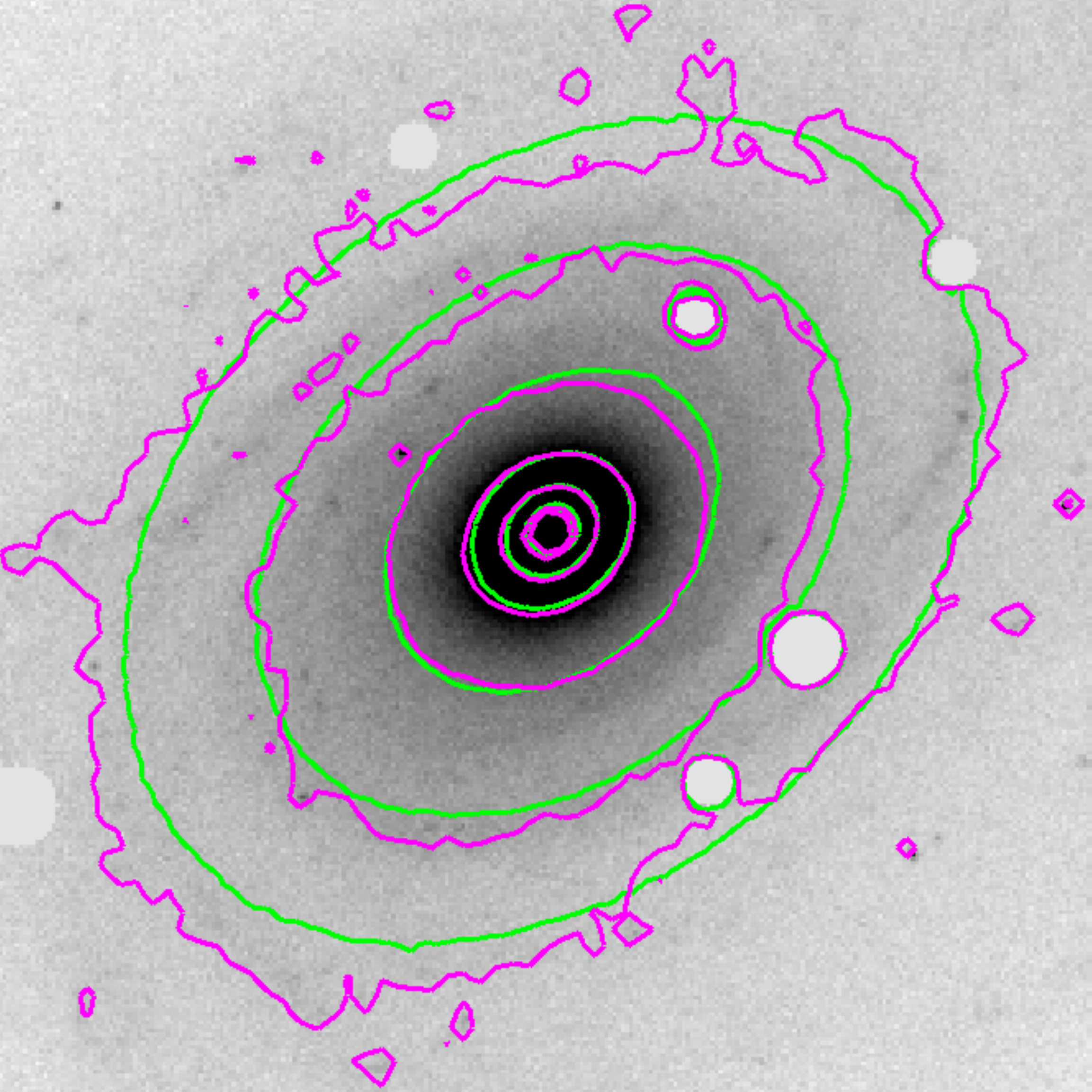}
\end{center}
\caption{
Isophotes of the Ks-band image of NGC~3223 (cyan contours) fitted with
FitSKIRT (green contours).  } 
\label{figure_fitskirt}
\end{figure}

The fitted isophotes with FitSKIRT are shown in Fig. \ref{figure_fitskirt}; the
best-fit effective radius of the bulge is 4 $\pm$ 0.5 arcsec with 
a S\'ersic index of $n=1.4 \pm 0.2$, and
the best-fit exponential scale-length of the disk is 18 $\pm$ 2
arcsec. We double-checked this value using 
{\sc BUDDA} and we found a consistent result.
The derived apparent Ks-band magnitudes of the disk and bulge
are 7.85$\pm$0.08 and 9.78$\pm$0.08, respectively.
The derived inclination angle (averaged over the five BVIHKs
bands) is 44.1$\pm 0.5^\circ$.

\subsection{Stellar kinematics}
\label{sec_stellar_kin}

One of the main goals of the present study is to determine the stellar velocity
ellipsoid, thus we have to take advantage of our observations of
stellar kinematics along the major and minor axes, mostly following
Gerssen 
\& Shapiro Griffin (2012, and references therein). Simple geometrical
considerations lead to the following relation between the observed
stellar velocity dispersions along the major and minor axes
($\sigma_{\rm maj}$ and $\sigma_{\rm min}$, respectively) 
and the three components of the velocity ellipsoid:

\begin{equation}
\sigma^2_{\rm maj}=\sigma^2_{\rm \theta} {\rm sin}^2 i +\sigma^2_{\rm z}
{\rm cos}^2 i 
\label{eq_sigmamaj}
\end{equation}

\begin{equation}
\sigma^2_{\rm min}=\sigma^2_{\rm R} {\rm sin}^2 i +\sigma^2_{\rm z} {\rm cos}^2 i 
\label{eq_sigmamin}
\end{equation}

where $\sigma_{\rm R}$,  $\sigma_{\rm  z}$ and $\sigma_{\rm \theta}$
are the radial, vertical, and tangential components of the velocity
dispersion, respectively. The inclination angle is indicated with
$i$. 

These are two equations and there are three unknowns, therefore a
third equation needs to be added. This is usually done by assuming
the epicycle approximation, which is valid in the disks of spiral
galaxies 
because the orbits of stars are close to circular:

\begin{equation}
\frac{\sigma^2_{\theta}}{\sigma^2_{\rm R}} = \frac{1}{2}
\left(1 + \frac{\partial {\rm ln} V_{\rm c}}{\partial {\rm ln} R} \right)
\label{eq_epicycle}
\end{equation}

where $V_{\rm c}$ is the circular velocity of the material in the disk
at a given galactocentric distance $r$. Equations \ref{eq_sigmamaj},
\ref{eq_sigmamin}, and \ref{eq_epicycle} 
represent a set of three equations and (in principle) three
unknowns. However, $V_{\rm c}$ is not necessarily exactly equal to the
measured tangential velocity $V_{\theta}$ because of asymmetric drift:

\begin{equation}
V^2_{\theta} = \frac{V^2_{\rm maj}}{{\rm sin^2}i} = V^2_{\rm c} -
\sigma^2_{\rm R} \left( \frac{R}{h} - R \frac{\partial}{\partial
    R} {\rm ln} \sigma^2_{\rm R} - \frac{1}{2} - \frac{R}{2 V_{\rm c}}  \frac{\partial V_{\rm c}}{\partial R} \right) + R\frac{\partial
    \sigma^2_{Rz}}{\partial z}
\label{eq_asymm}
\end{equation}

where $h$ is the exponential scale length of the stellar
disk and $\sigma^2_{Rz}$ is the tilt term, which can be safely
neglected in these observations of external galaxies (Gerssen et
al. 1997, 2000). 

If no gas kinematics are present, the four equations
\ref{eq_sigmamaj}, \ref{eq_sigmamin}, \ref{eq_epicycle}, and
\ref{eq_asymm} are used to determine the four unknowns $\sigma_{\rm
  R}$,  $\sigma_{\rm _z}$ and $\sigma_{\rm \theta}$, and $V_{\rm
  c}$. In our case, we have two gas kinematics observations: 2D
H$\alpha$ kinematics from Palunas \& Williams (2000) and our own 
H{\sc i} data (radially more extended but with a lower angular
resolution). We will use the H$\alpha$ data as an extra piece of information and
as a consistency check for the
circular velocities determined from stellar kinematics.

However, direct inversion of the equations to determine the unknowns
as a function of radius results in very large uncertainties due to
error propagation and quite noisy 
data to start with. 

Therefore we followed the approach taken
by e.g. Gerssen \& Shapiro Griffin (2012, and references therein), where the
radial and vertical components of the velocity dispersion are modelled 
as exponentials (in the radial direction):

\begin{equation}
\sigma_{\rm R} = \sigma_{\rm R,0} e^{-R/a_{\rm R}}
\label{eq_sigmar}
\end{equation}

\begin{equation}
\sigma_{\rm z} = \sigma_{\rm z,0} e^{-R/a_{\rm z}}
\label{eq_sigmaz}
\end{equation}

and the circular velocity as a function of radius (the rotation curve)
is modelled as a power-law:

\begin{equation}
V_{\rm c} = V_{0} R^\alpha
\label{eq_vc}
\end{equation}

Hence the problem reduces to modelling the observed 
$\sigma_{\rm maj} (r)$, $\sigma_{\rm min}(r)$, $V_{\rm maj}(r)$,
and gas rotation curve with 
six parameters ($\sigma_{\rm R,0}$, $\sigma_{\rm z,0}$, 
$a_{\rm R}$, $a_{\rm z}$, $V_{0}$, and $\alpha$).

The central parts of the galaxy are clearly dominated by the
bulge. Therefore, in our analysis we only consider the radii where 
the disk-to-bulge brightness ratio is larger than 20. Based on the
disk and bulge parameters derived in section \ref{sec_photometry}, we thus only
consider radii larger than 3.2 kpc (17.3 arcsec).

In a locally isothermal disk
the vertical density distribution is given by sech$^2(z/z_0)$ where
$z$ is the vertical distance from midplane and $z_0$ is the vertical
scale-height. In this case, the disk surface density $\Sigma$, 
and the scale-height $z_0$ are linked to the vertical velocity
dispersion through the following relation:

\begin{equation}
\sigma_z^2 = \pi G \Sigma z_0 
\label{eq_disk_mass}
\end{equation}

As shown by e.g. Bottema (1993), if the disk mass-to-light (M/L) ratio
is constant with radius, then  the equation 
above can be rewritten as follows: 

\begin{equation}
\sigma_z^2 = \pi G (M/L) I_{K_s} z_0
\label{eq_disk_mass2}
\end{equation}

and $I_{K_s}=I_0 e^{-R/h}$ is the surface brightness of the stellar disk (in the
Ks-band) measured here. 
Assuming the kinematic scale-length (for the vertical velocity dispersion)
is twice the photometric scale-length, i.e. $a_{\rm z}=2h$ (Bottema
1993, Binney \& Merrifield 1998), we can rewrite
eq. (\ref{eq_disk_mass2}) to get an independent estimate of the 
M/L ratio:

\begin{equation}
(M/L) = \frac{\sigma_{\rm z,0}^2}{\pi G I_0 z_0}
\label{eq_disk_mass3}
\end{equation}

In our analysis, we will investigate models where $a_{\rm z}=2h$,
$a_{\rm z}=h$, and $a_{\rm z}$ is a free parameter.

\begin{figure}
\begin{center}
\includegraphics[width=0.47\textwidth]{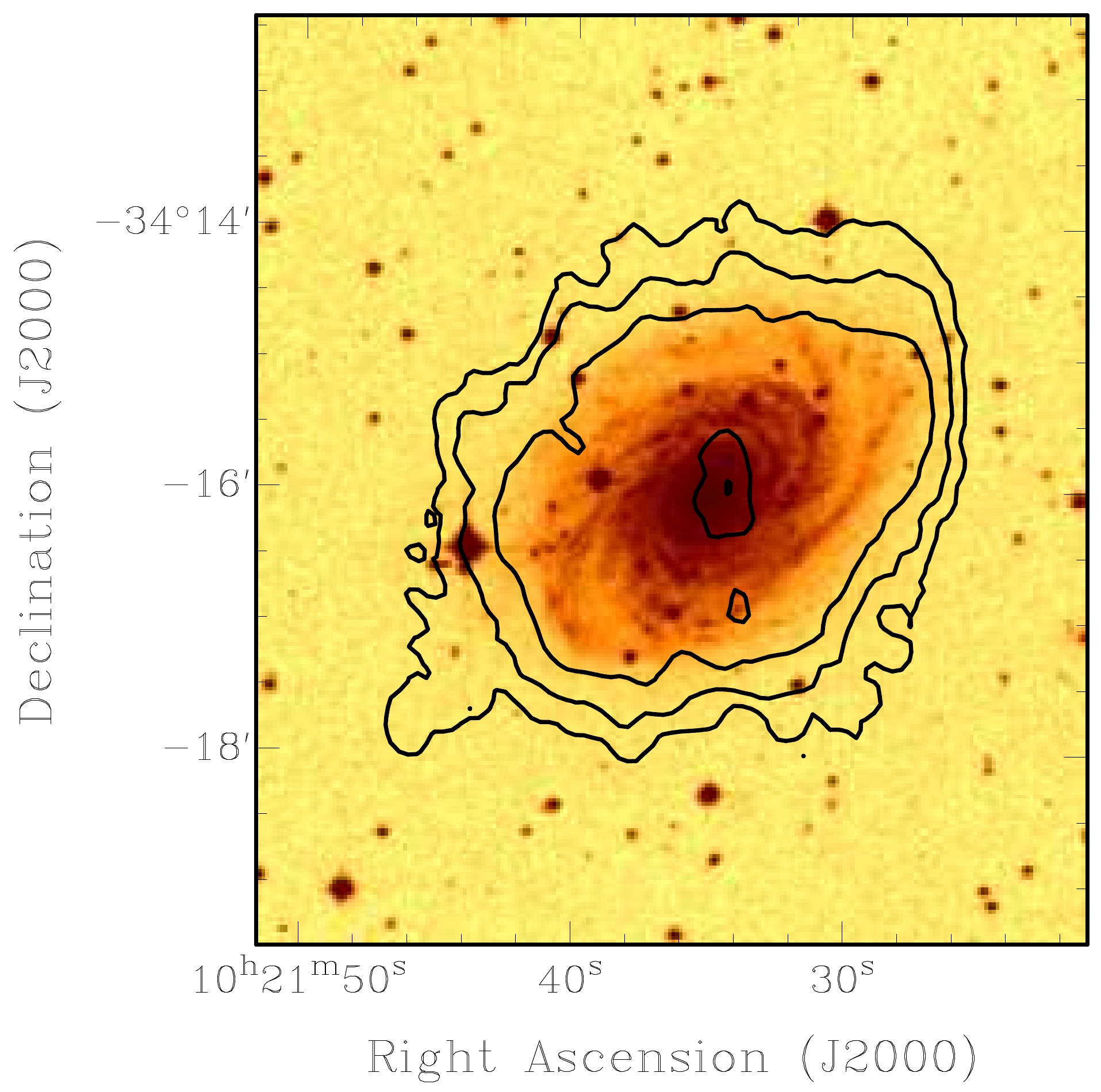}
\end{center}
\caption{
Total H{\sc i} map of NGC~3223 (contours) overlaid onto an optical DSS image
(colour scale). Contour levels are (1.5, 3, 6, 12) $\times 10^{20}$ atoms
cm$^{-2}$. Note that the central contour represents a drop in H{\sc i}
surface density.} 
\label{m0_opt}
\end{figure}

\begin{figure}
\begin{center}
\includegraphics[width=0.49\textwidth]{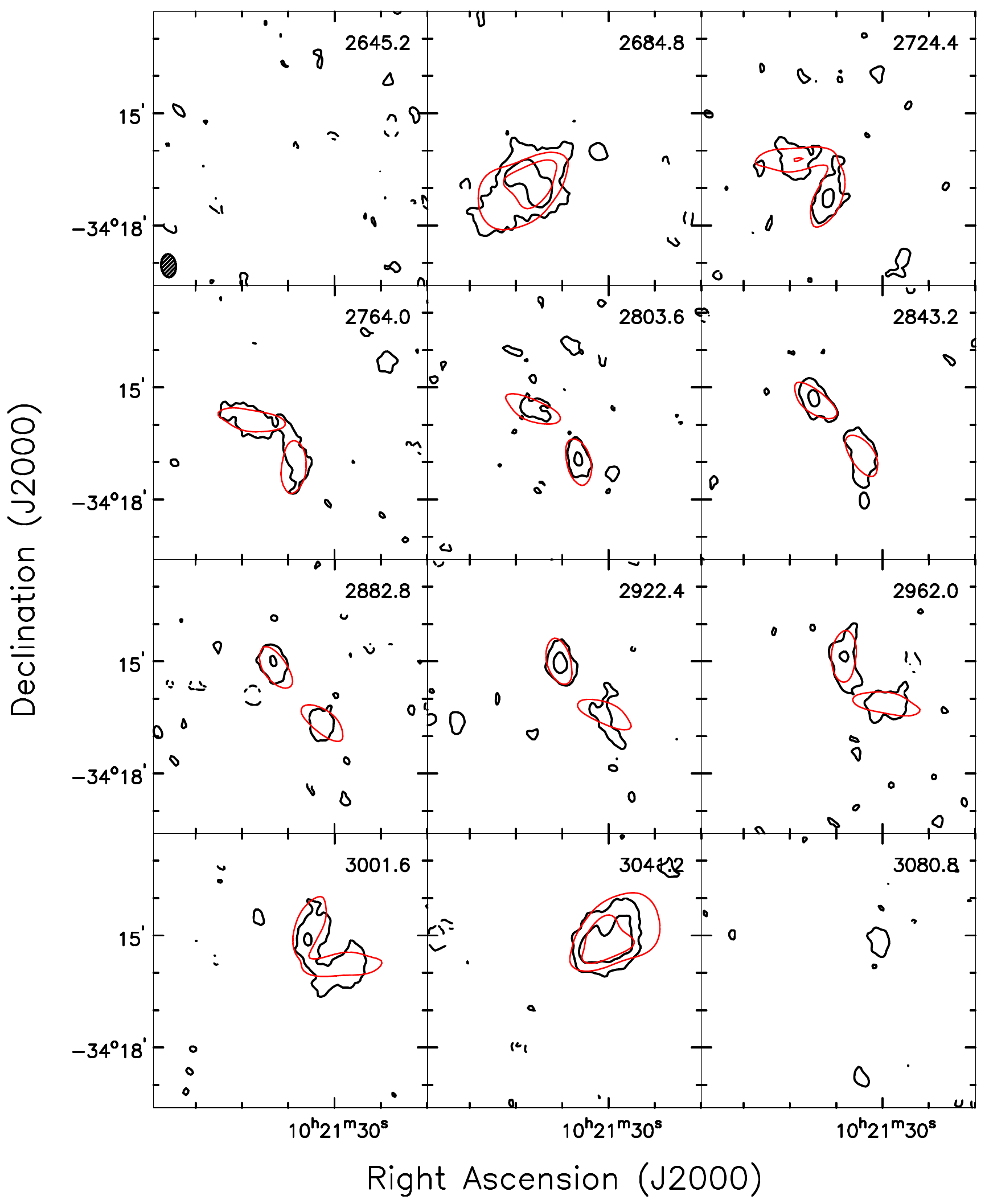}
\end{center}
\caption{
H{\sc i} data cube of NGC~3223. The beam is shown
in the bottom left of the top left panel (36.9 $\times$ 23.6 arcsec). The contours are -3, 3
(2.5$\sigma$), and 9 mJy beam$^{-1}$. Negative contours
are dashed. The observed data cube is shown as thick black contours
and the model presented here is shown as thin red contours. } 
\label{figure_cube}
\end{figure}

\begin{figure}
\begin{center}
\includegraphics[width=0.47\textwidth]{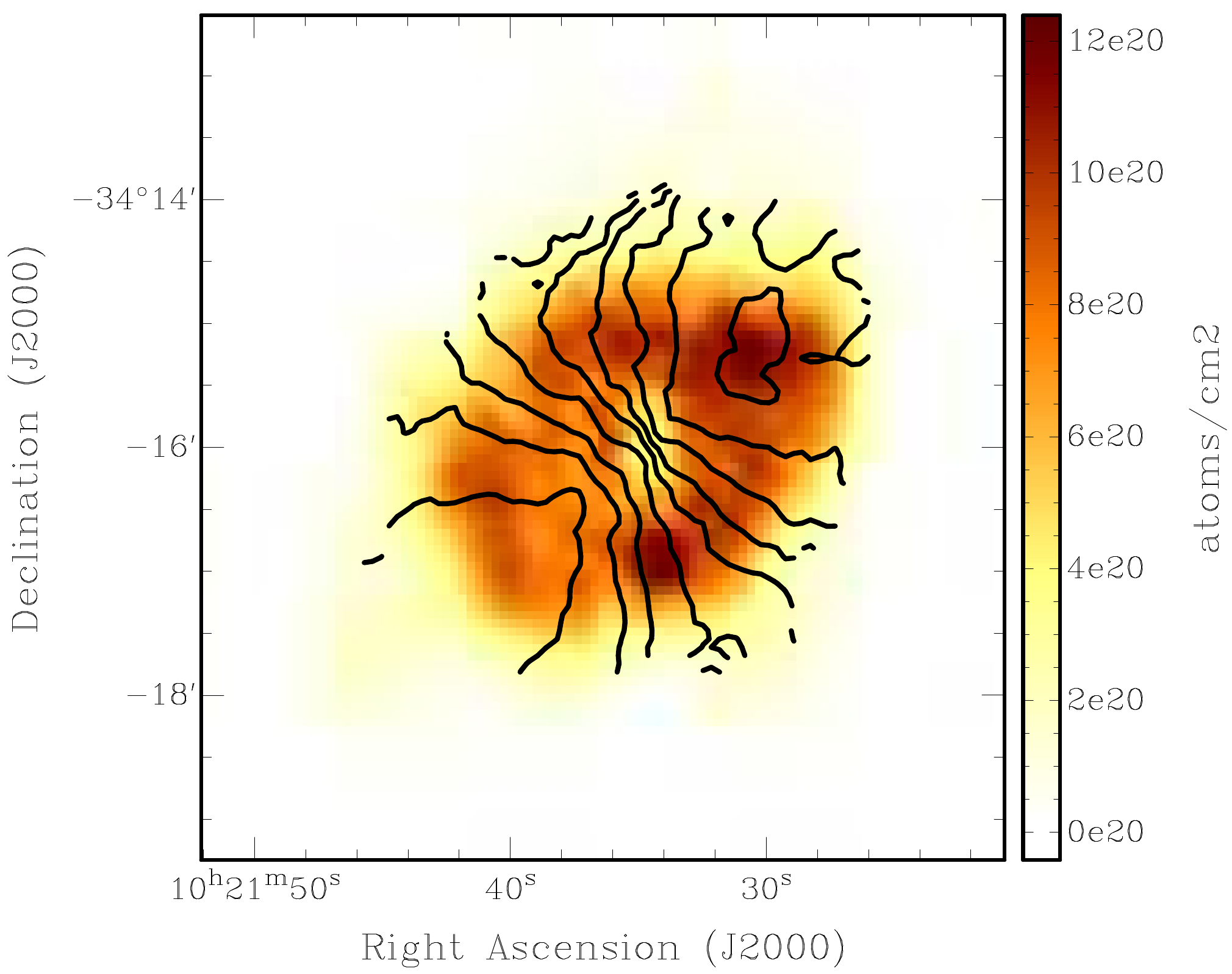}
\end{center}
\caption{
Intensity-weighted H{\sc i}  velocity field of NGC~3223 (contours)
overlaid onto the total H{\sc i} map
(colour scale). Contour levels start from 2704 km s$^{-1}$ in steps of
40 km s$^{-1}$. } 
\label{m0_velfi}
\end{figure}

\subsection{H{\sc i} data cube analysis}

Using the primary beam-corrected data cube, we find a total H{\sc i} flux of
26.6 Jy km s$^{-1}$, corresponding (at a distance of 38.1 Mpc) to 9.1
$\times$ 10$^9$ M$_\odot$.
The total H{\sc i} map is superimposed
onto an optical image in Fig. \ref{m0_opt}.

Thorough modelling of the data cube of NGC~3223 goes beyond the aim of
this paper, but we made some simple models to have a good
understanding of the H{\sc i} distribution and kinematics. We used the code
called TiRiFiC (J\'ozsa et al. 2007) to make models of the H{\sc i} data
cube, which we iteratively compared to the observed one
(see Gentile et al. 2013 for more details about the procedure). The best-fit
model is shown in Fig. \ref{figure_cube}: it gives a very good
representation of the observations. The data cube, Fig. \ref{figure_cube} and the
total H{\sc i} map indicate that NGC~3223 has a regularly rotating
H{\sc i} disk, see the velocity field shown in Fig. \ref{m0_velfi}.
The gaseous disk also has an extension to the SE around position (10 21 45,
-34 17 30), see Fig. \ref{m0_opt}. This extension is present in the
channel maps between velocities 2678 and 2704 km s$^{-1}$.

Particular attention was given to the derivation of the rotation
curve. A first estimate was derived from the tilted-ring fit of the
intensity-weighted velocity field, then the rotation curve was
iteratively modified to match the observed data cube, in particular
the position-velocity diagram along the major axis. Each data point
is assigned an uncertainty that is the maximum between two quantities:
(1) the difference
between the two sides (receding and approaching), and (2) the velocity
resolution corrected for inclination. The best-fit inclination in
the inner parts (52$^\circ$, but with an uncertainty of about
10$^\circ$ because of the relatively irregular shape of the H{\sc i}
contours) 
is slightly higher than the one derived
from the optical and NIR data (44.1$^\circ$, see section \ref{sec_photometry}). When
plotting the two rotation curves together, the H{\sc i} data points were
corrected for the inclination difference; we consider the inclination
from the optical and NIR data to be more reliable because of the less
regular shape of the H{\sc i} emission.

\begin{table*}
\caption{Best-fit parameters of the joint $\sigma_{\rm min}$ -
  $\sigma_{\rm maj}$ - $V_{\rm rot}$ - $V_{\rm c}$ fit. The stellar
  exponential scale-length $h$ is  18 $\pm$ 2
 arcsec, corresponding to 3.3 $\pm$ 0.4 kpc.}              

\label{tab-fit}      
\centering                                     
\begin{tabular}{l l l l l l l}  
\hline\hline     
Fit                                                    &   $\sigma_{\rm R,0}$   & $\sigma_{\rm z,0}$ &$a_{\rm  R}$            & $a_{\rm z}$            & $V_{0}$                   &            $\alpha$       \\    
                                                        &   km s$^{-1}$           &   km s$^{-1}$        &kpc                      &          kpc             &   km s$^{-1}$          &                                   \\    
\hline
All free                                             & 182$\pm$29               & 140$\pm$16  & 5.1$\pm$1.3       & 12.7$\pm$3.4     & 280.6$\pm$1.5       &0.064$\pm$0.009    \\
$a_{\rm R}$=$a_{\rm  z}$                      & 135$\pm$12         & 162$\pm$12        & 8.9$\pm$0.8       & =$a_{\rm  R}$          & 284.5$\pm$1.8     &0.097$\pm$0.014   \\
$a_{\rm  z}$=2$h$                              & 129$\pm$15     & 192$\pm$11    & 10.2$\pm$1.8        &6.6 (fixed)             & 285.9$\pm$1.9     &0.102$\pm$0.014    \\
$a_{\rm  z}$=2$a_{\rm R}$=2$h$          & 232$\pm$13     &  222$\pm$8  & 3.3 (fixed)         &6.6 (fixed)                     & 275.8$\pm$1.2        &0.053$\pm$0.007    \\
\hline       
\end{tabular}
\end{table*}

\section{Results}
\label{sec_results}

The first fit we made was with the six parameters ($\sigma_{\rm R,0}$, $\sigma_{\rm z,0}$, 
$a_{\rm R}$, $a_{\rm z}$, $V_{0}$, and $\alpha$) described in Section
\ref{sec_stellar_kin}, fitting simultaneously the major axis stellar velocity
dispersion profile, the minor axis stellar velocity dispersion profile, the
major axis stellar rotation velocity profile, and the circular
velocity curve as derived from the gas kinematics (H$\alpha$, the H{\sc i} data were
  used for the mass decomposition fits but not for the combined
  velocity dispersion/circular velocity fits).
The fit quality is very good but some parameters are rather weakly
constrained. The results of all the fits are tabulated in Table
\ref{tab-fit}, and the fits are plotted in Fig. \ref{fit_dec28}.

To investigate the real uncertainties on the parameters in a better
way, we then made a fit where we forced the two kinematic scale-lengths
(radial and vertical,  $a_{\rm  r}$ and $a_{\rm  z}$ respectively) to
have the same value, as in Shapiro et al. (2003) and Gerssen \& Shapiro
Griffin (2012).
The next choice was to impose $a_{\rm  z}$=2$h$ in order to be able to
use eq. (\ref{eq_disk_mass3}) to estimate the stellar M/L ratio. Lastly,
we made a fit with only four free parameters ($\sigma_{\rm R,0}$, $\sigma_{\rm z,0}$, 
$V_{0}$, and $\alpha$) and fixing $a_{\rm R}=h$ (to have a constant Toomre
stability parameter with radius) and $a_{\rm z}=2h$ (like above, to
obtain an estimate of the stellar M/L ratio).  The uncertainties on the fitted
parameters were derived by creating 100 randomly generated data sets from Gaussian
distributions defined by the observed data. 
This means for each realisation we generated a new data point,
corresponding to each original data point, by randomly sampling from
the original point with a Gaussian distribution with variance equal to
our originally computed variance. These new data points were assumed
to have the same variance as the originals. 
These randomly generated
data were then subjected to the same fitting procedure ($\chi^2$
minimisation) as the original
data, which gives distributions of values for the best-fit
parameters and hence estimates of their 1-$\sigma$ uncertainties from
Gaussian fits of their distribution. 

\subsection{Disk heating}

\begin{figure*}
\begin{center}
\includegraphics[width=0.9\textwidth]{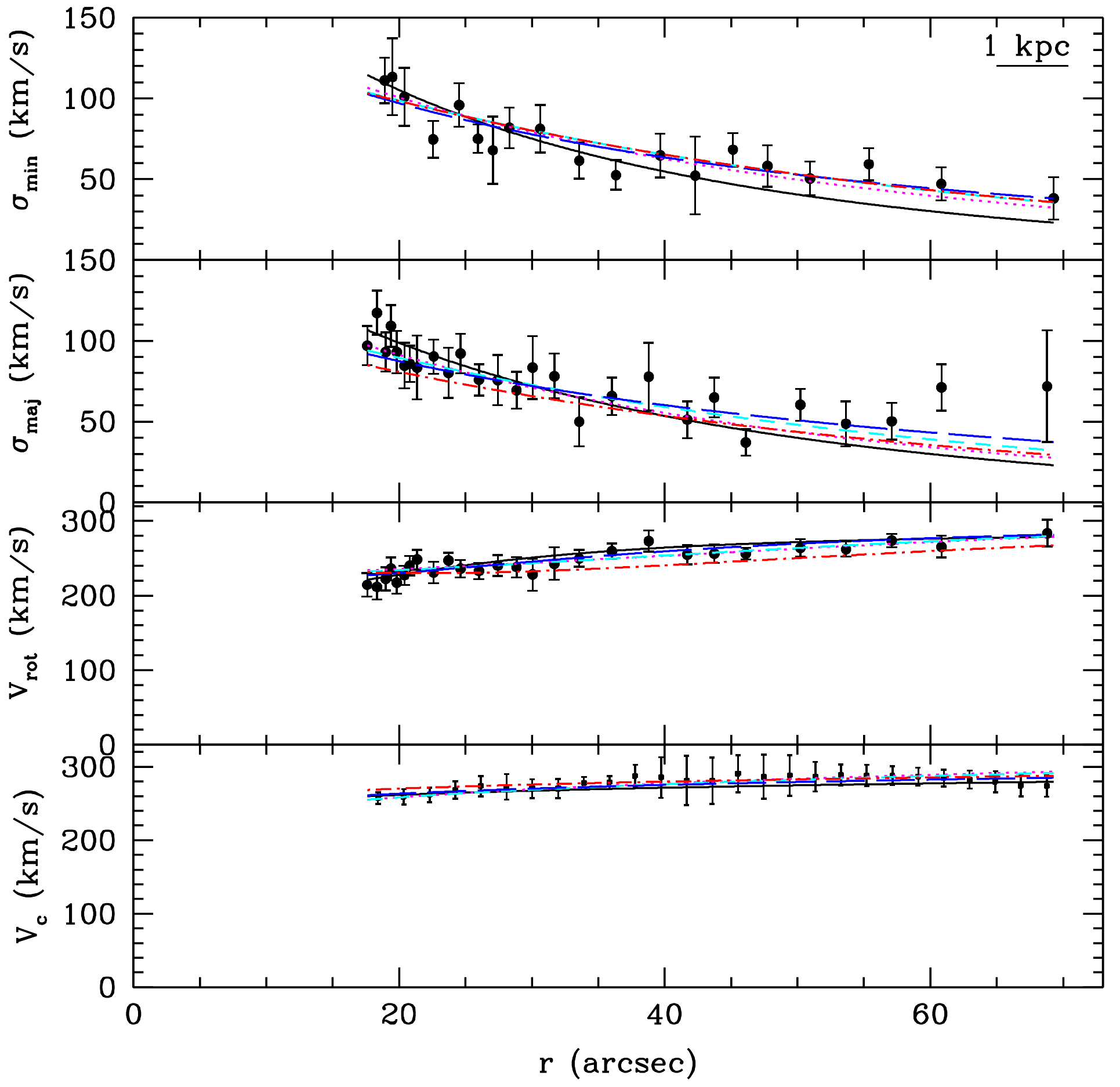}
\end{center}
\caption{
Fit with all six free parameters (blue long-dashed lines), fit with $a_{\rm
  z}$=$a_{\rm R}$ left as a free parameter (cyan short-dashed lines), fit with
$a_{\rm z}$ fixed at 6.6 kpc (twice the disk scale-length) and 
$a_{\rm R}$ free (magenta dotted lines), fit with $a_{\rm z}$ fixed at
6.6 kpc (twice the disk scale-length) and 
$a_{\rm R}$ fixed at the disk scale-length, 3.3 kpc (black solid lines), and
fit it with $a_{\rm z}$=$a_{\rm R}$ left as a free parameter, and with
$\sigma_{\rm z}/\sigma_{\rm R}$ fixed at 0.63 (red dot-dashed lines).
Note that the bottom panel has a  
different horizontal scale from the top three panels. In the bottom
panel the H$\alpha$ data (full circles) come from Palunas \& Williams
(2000).} 
\label{fit_dec28}
\end{figure*}

From observations of eight spiral galaxies,
Gerssen \& Shapiro Griffin (2012) find a relation between Hubble type and
$\sigma_{\rm z}/\sigma_{\rm R}$ ratios: it turns out that later type galaxies have
lower $\sigma_{\rm z}/\sigma_{\rm R}$ ratios. From the relation they derive, the
expected value 
for an Sb galaxy like NGC~3223 
would be around 0.63 with an uncertainty of about 0.2. 
From the fit with $a_{\rm  r}=a_{\rm  z}$
we find
$\sigma_{\rm z}/\sigma_{\rm R}=1.21 \pm 0.14$, not consistent with
Gerssen's relation. If we force the ratio to be equal to 0.63, we find
(Fig. \ref{fit_dec28}) that the fit indeed worsens, in
particular the major axis velocity dispersion and rotation velocity
profiles are underestimated. 

It is possible that we overestimated the inclination
(44.1$^\circ$). By taking 40.5$^\circ$, the value of Kassin et
al. (2006a), we find a best-fit ratio of $\sigma_{\rm z}/\sigma_{\rm R}=1.28
\pm 0.17$, still higher than the value suggested by Gerssen \& Shapiro
Griffin (2012). 

It is also possible that the choice of cutoff radius is incorrect. 
Changing the inner cutoff radius to 15 arcsec (instead
of 17.3 arcsec) gives an almost identical value, $\sigma_{\rm
  z}/\sigma_{\rm R}=1.19\pm0.11$. Changing it to 25 arcsec gives
$\sigma_{\rm z}/\sigma_{\rm 
  R}=1.07\pm0.19$, slightly lower but (as intuitively expected) with a
larger uncertainty, because there are fewer data points to constrain
the parameters. 

The result that NGC 3223 is more vertically heated
than expected implies that NGC 3223 is an outlier in the observational relation of
Gerssen \& Shapiro Griffin (2012). The reason for this large vertical
heating is not obvious, but recent mergers are excluded because of the
regular morphology and kinematics.
We note that the values we found for NGC~3223 are marginally inconsistent with the
correlation between $\sigma_{\rm z}$ and $\sigma_{\rm z}/\sigma_{\rm R}$ found by
Gerssen \& Shapiro Griffin (2012). They plot the velocity dispersions
at half the kinematic scale-length. With
$\sigma_{\rm z}=98.3\pm7.3$ km s$^{-1}$ and
$\sigma_{\rm z}/\sigma_{\rm R}=1.21 \pm 0.14$, 
NGC~3223 is (marginally) inconsistent with the relation found in their
Fig. 6.

\subsection{Constraint on the M/L ratio}
\label{sec_ml}

The evaluation of constraints on the stellar M/L was done using the
results of the fits made with the scale length of the vertical
velocity dispersion distribution fixed at twice the stellar
exponential scale length (see Section \ref{sec_results}).

Applying eq. (\ref{eq_disk_mass3}) to the fits using 
$a_{\rm
  z}$=2$h$ and $a_{\rm
  z}$=2$a_{\rm R}$=2$h$ in Table \ref{tab-fit} gives values of
M/L=0.96$\pm$0.34 and 1.28$\pm$0.46, respectively, 
which is quite high for the Ks-band. The errors quoted here are formal uncertainties based on the fits
and error propagation. We used a sech$^2$ vertical distribution with a
scale-height of 0.9 kpc,
based on the Ks-band exponential scale-length and using the average between 
the scale-length-to-scale-height ratios given by Kregel et al. (2002)
and Bershady et al. (2010b). 
We used the Ks-band absolute magnitude of the Sun given by Pecaut \&
Mamajek (2013), i.e. 3.30.

Could other effects contribute to the real
uncertainties? One possibility is the inclination angle: if we assume
40.5$^\circ$ (see previous section) the results above become
0.91$\pm$0.33 and 1.08$\pm$0.39, respectively. 
Changing the inner cutoff radius to 15 arcsec instead
of 17.3 arcsec gives (for $a_{\rm  z}$=2$h$ and $a_{\rm  z}$=2$a_{\rm
  R}$=2$h$) 0.98$\pm$0.35 and 1.25$\pm$0.45, thus a negligible
variation in the derived stellar M/L ratio. Changing the inner cutoff
radius to 25 arcsec instead of 17.3 arcsec gives 0.75$\pm$0.27 and 
and 1.36$\pm$0.49, respectively.

The expectations from stellar
population synthesis range from 0.35 to 0.80, using a B-V colour of
0.80 and the relations between colour and stellar M/L ratio found in
Bell \& de Jong (2001), Bell et al. (2003), Portinari et al. (2004),
Zibetti et al. (2009), and Into and Portinari (2013). The B-V colour
seems to be a good indicator of M/L ratio (McGaugh 
\& Schombert 2014), even though at NIR wavelengths the dependency of
M/L on colour is weak.

Therefore we can conclude that our estimated
range of stellar M/L ratios (between 0.5 and 1.7) is consistent with the
most massive predictions of stellar population synthesis models. 
We have checked that within the outermost point fitted in the velocity
dispersion profiles the stellar surface density (even with a stellar
M/L ratio of 0.5) is much higher ($>$ 10 times) than the H{\sc i} surface
density: thus our results about the M/L ratio are not falsified by the
H{\sc i} surface density influencing the vertical velocity dispersion.

\subsection{Rotation curve fitting}

\begin{figure*}
\centering
\begin{minipage}[b]{0.32\textwidth}
\includegraphics[width=\textwidth]{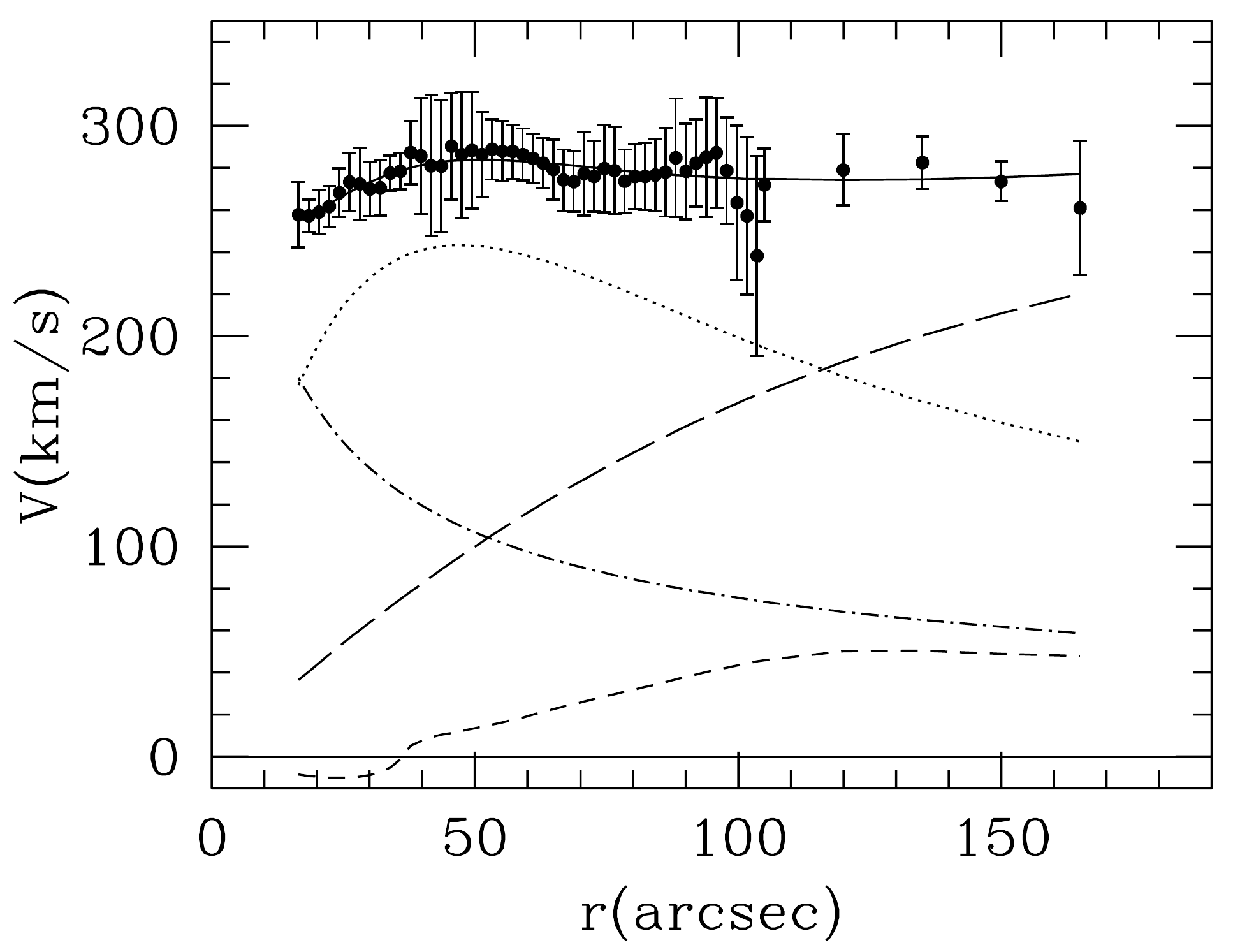}
\end{minipage}
\hspace{0.01em}
\begin{minipage}[b]{0.32\textwidth}
\includegraphics[width=\textwidth]{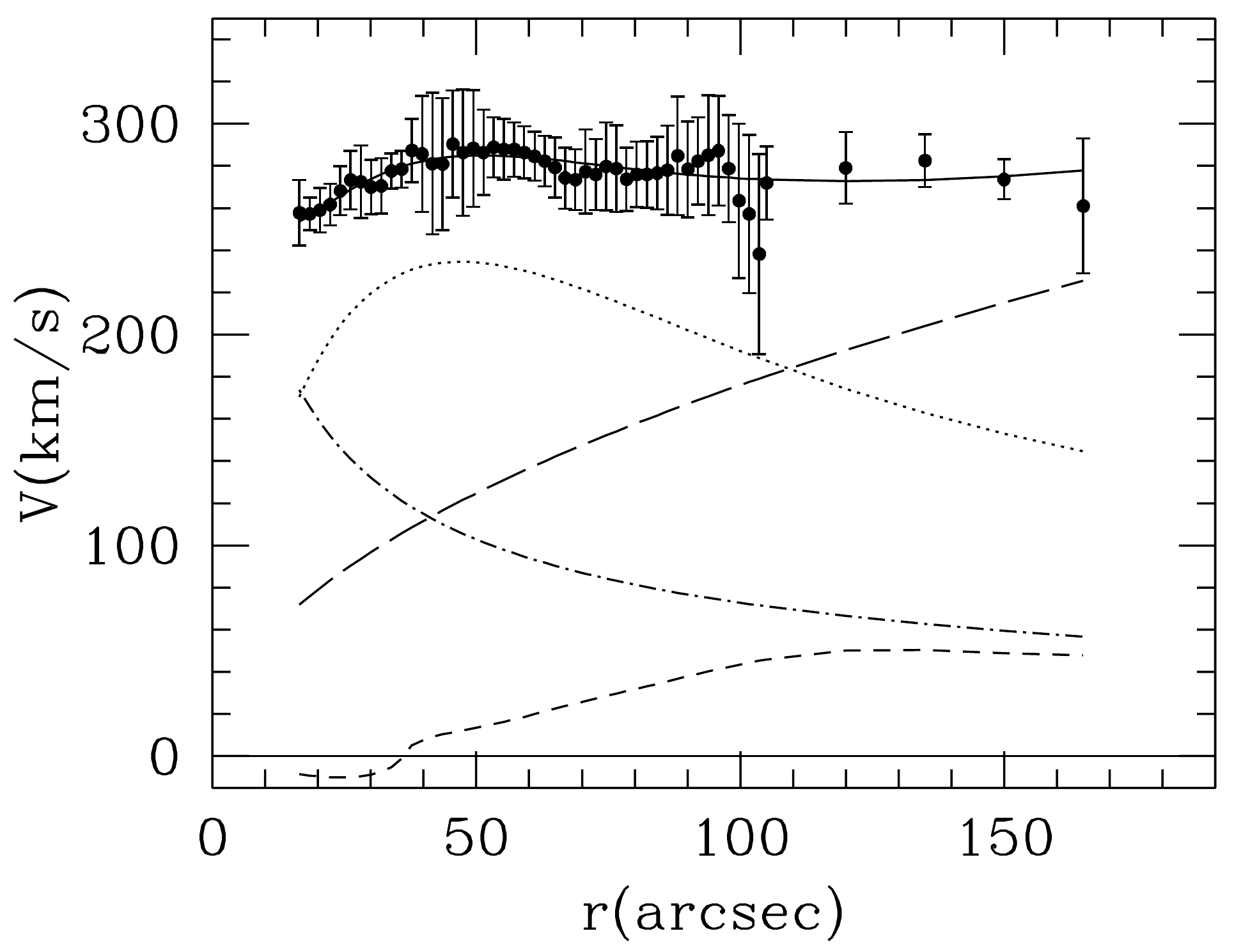}
\end{minipage}
\hspace{0.01em}
\begin{minipage}[b]{0.32\textwidth}
\includegraphics[width=\textwidth]{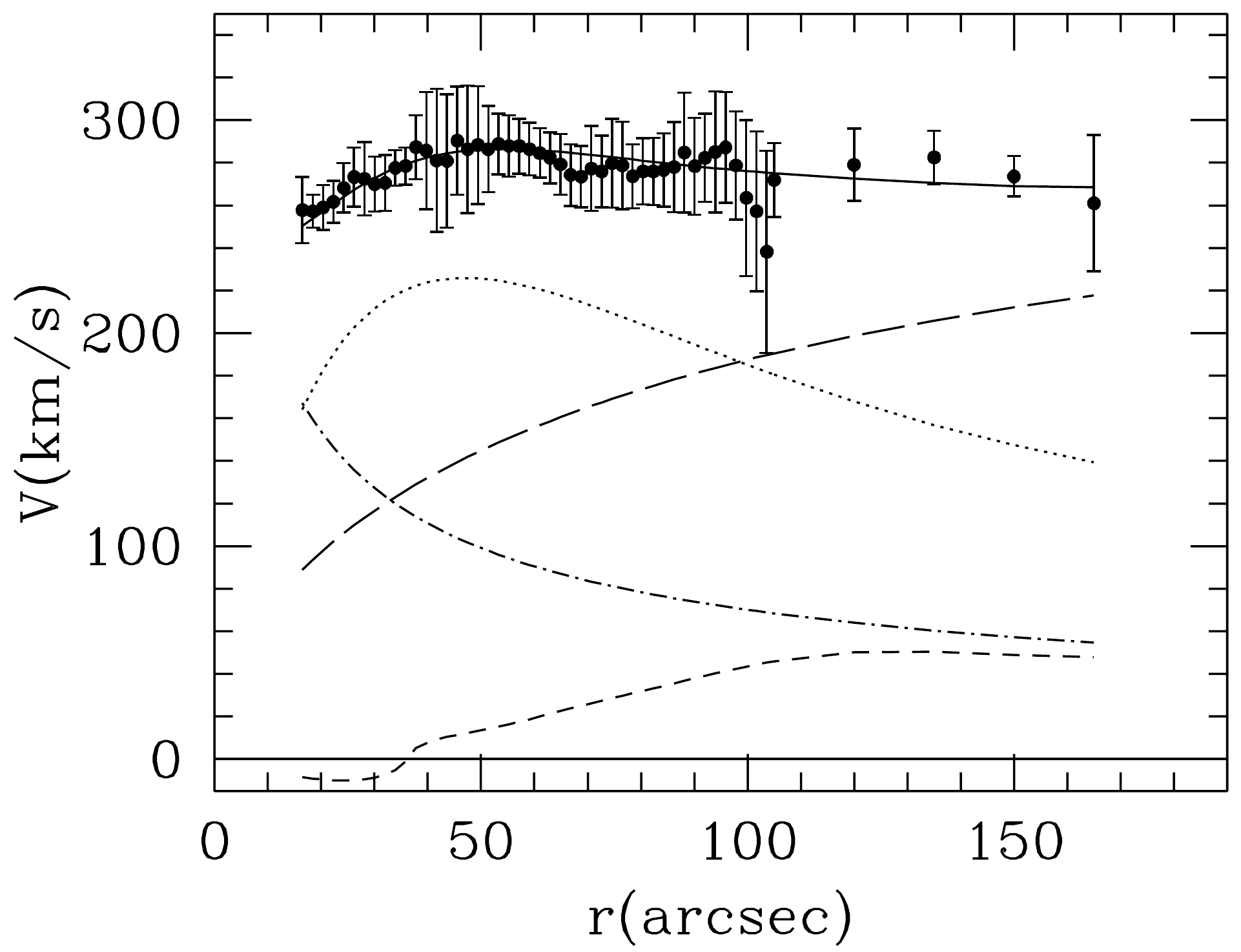}
\end{minipage}
\caption{Rotation curve fits. Left panel: Burkert halo, middle:
  unconstrained NFW halo; right panel: NFW halo using the $c-M_{\rm
    vir}$ relation. The points with errorbars represent the combined
  H$\alpha$+H{\sc i} rotation curve. The short dashed, long
  dashed, dotted, and dot-dashed curves represent the gas, halo,
  stellar disk, and bulge contributions, respectively. The total
  (best) fit to the rotation curve for a particular model is
  represented by a solid line.}
\label{rcfit}
\end{figure*}

\begin{table*}
\caption{Best-fit parameters of the rotation curve fits. For the
  Burkert halo ``Parameter 1''
  is the core radius in kpc and ``Parameter 2'' is the central density
in M$_{\odot}$ pc$^{-3}$, whereas for the NFW halo ``Parameter 1'' is
the virial mass in 10$^{11}$ M$_{\odot}$ and ``Parameter 2'' is the
concentration.}

\label{tab-fit2}      
\centering                                     
\begin{tabular}{l l l l l l l}  
\hline\hline      
Fit                                                      &   Stellar M/L$_{K}$ ratio  & Parameter 1  & Parameter 2  \\    
\hline
Burkert halo                                       & 0.71$^{+0.04}_{-0.04}$      &  26.9 $^{+31.5}_{-10.5}$  &0.0087$^{+0.0075}_{-0.0040}$      \\
Unconstrained NFW halo                    & 0.66$^{+0.03}_{-0.08}$     & 10545$^{+5530}_{-10510}$  & 1$^{+9.6}_{-0}$      \\
NFW halo with c-M$_{\rm vir}$ relation & 0.61$^{+0.02}_{-0.02}$     &  57$^{+22}_{-18}$  & 8.0$^{+0.5}_{-0.4}$  \\
\hline       
\end{tabular}
\end{table*}

Using the constraints on the stellar M/L ratio derived in Section
\ref{sec_ml} we made rotation curve fits using different dark matter
halo density distributions. The bulge and disk contributions to the
rotation curve were derived from the parameters of the Ks-band bulge/disk
decomposition of Section \ref{sec_photometry}. 
For the bulge we adopt the same stellar M/L ratio as for the disk. The H{\sc i} disk
contribution was derived from the observed surface density
distribution (corrected for primordial He), and for the dark matter
halo we used two different functional forms:

$\bullet$ the Burkert (1995) halo,
known to give good fits to rotation curves (e.g. Salucci et al. 2007).
The density $\rho(R)$ as a function of radius $R$ has a central core
(i.e. a finite central density) and is given by:  

\begin{equation}   
\rho_{\rm Bur}(R)=\frac{\rho_0 R_{\rm core}^3}{(R+R_{\rm core})   
(R^2+R_{\rm core}^2)}   
\end{equation}   
   
where the central density $\rho_0$ and the core radius $R_{\rm core}$ are    
the two free parameters.    

$\bullet$ the NFW halo (Navarro, Frenk \& White 1996), characterised
by a central density cusp, whose functional form is the following:

\begin{equation}   
\rho_{\rm NFW}(R)=\frac{\rho_{\rm s}}{(R/R_{\rm s})(1+R/R_{\rm s})^2},    
\end{equation}   
   
\noindent    
where $\rho_{\rm s}$ and $R_{\rm s}$ are the characteristic density and    
scale-length of the dark matter halo. The parameters $\rho_{\rm s}$
and $R_{\rm s}$ can also be expressed in terms of the virial mass
$M_{\rm vir}$ and the concentration $c=R_{\rm vir}/R_{s}$ ($R_{\rm
  vir}$ is the virial radius). Dark matter simulations of structure
formation in the Universe show that $c$ and $M_{\rm vir}$ are not
independent (e.g. Neto et al. 2007). We adopt here the $c-M_{\rm vir}$
relation used by Gentile et al. (2007):

\begin{equation}
c_{\rm vir} \simeq 13.6 \left( \frac{M_{vir}}{10^{11}{\rm M}_{\odot}} \right)^{-0.13}
\end{equation}

The fits are plotted in Fig. \ref{rcfit}: the Burkert halo fit is very
good, and the product $\rho_0 R_{\rm core}=232^{+338}_{-133}$
M$_\odot$ pc$^{-2}$ is consistent with the value found by Donato et
al. (2009) and Gentile et al. (2009) for a large sample of galaxies.

The NFW halo fits are almost as good. In the fit with no constraints
on the virial mass and concentration the allowed range of virial
masses spans almost
three orders of magnitude. This range would have been larger if we had
not put a minimum reasonable value of 1 for the concentration
parameter. The constrained NFW halo fit is only slightly worse
($\chi^2$ value $\sim$ 35\% higher), we
find a value of the virial mass of 5.6$^{+2.1}_{-1.7} \times 10^{12}$
M$_{\odot}$, consistent with the value of the unconstrained fit.

\section{Conclusions}
\label{sec_conclusions}

We have presented observations of the stars and gas kinematics of
NGC~3223, an Sb galaxy located at a 
distance of about 38 Mpc. 

From the observed stellar velocity dispersion along the major and
minor axis, from the major-axis stellar rotation velocity and from the
gas kinematics, we determined the vertical and radial stellar
velocity dispersion as a function of radius. We found a vertical-to-radial velocity dispersion ratio of
$\sigma_{\rm z}/\sigma_{\rm R}=1.21 \pm 0.14$, significantly higher 
the relation between $\sigma_{\rm z}/\sigma_{\rm R}$ and Hubble type found by
e.g. Gerssen \& Shapiro Griffin (2012).

We also used the Ks-band photometry together with the derived vertical
velocity dispersion to estimate the mass-to-light (M/L) ratio
of the stellar disk. We find that the uncertainties are quite large,
and the allowed range lies between 0.5 and 1.7 (in the Ks-band). This lies
at the high end of the values predicted by stellar population
synthesis models. Because of such a weak constraint on the stellar M/L
ratio we cannot really constrain the distribution of dark matter in
NGC~3223. However, the method proved to yield meaningful results, consistent
with other data, therefore a statistical approach involving a sample
of galaxies with similar data and a spread in physical properties will certainly
provide more answers.

\section*{Acknowledgements}

We thank the anonymous referee for constructive comments that improved
the content and presentation of this paper. 
The research of GWA supported by the FWO - Vlaanderen. MK is a
postdoctoral fellow of the Fund for Scientific Research- Flanders,
Belgium (FWO 65052 / 12E4115N LV).


\begin{thebibliography}{}

\bibitem[Appenzeller et al.(1998)]{1998Msngr..94....1A} Appenzeller, I., 
Fricke, K., F{\"u}rtig, W., et al.\ 1998, The Messenger, 94, 1 

\bibitem[Bell 
\& de Jong(2001)]{2001ApJ...550..212B} Bell, E.~F., \& de Jong, R.~S.\ 2001, \apj, 550, 212 

\bibitem[Bell et al.(2003)]{2003ApJS..149..289B} Bell, E.~F., McIntosh, 
D.~H., Katz, N., \& Weinberg, M.~D.\ 2003, \apjs, 149, 289 

\bibitem[Bershady et al.(2010)]{2010ApJ...716..198B} Bershady, M.~A., 
Verheijen, M.~A.~W., Swaters, R.~A., et al.\ 2010a, \apj, 716, 198 

\bibitem[Bershady et al.(2010)]{2010ApJ...716..234B} Bershady, M.~A., 
Verheijen, M.~A.~W., Westfall, K.~B., et al.\ 2010b, \apj, 716, 234 

\bibitem[Bevington 
\& Robinson(2003)]{2003drea.book.....B} Bevington, P.~R., \& Robinson,
D.~K.\ 2003, Data reduction and error analysis for the physical
sciences, 3rd ed., by Philip R.~Bevington, and Keith
D.~Robinson.~Boston, MA: McGraw-Hill, ISBN 0-07-247227-8, 2003

\bibitem[Binney 
\& Merrifield(1998)]{1998gaas.book.....B} Binney, J., \& Merrifield,
M.\ 1998, Galactic astronomy / James Binney and Michael Merrifield.~
Princeton, NJ : Princeton University Press, 1998.~ (Princeton series
in astrophysics) QB857 .B522 1998 

\bibitem[Bonnarel et 
al.(2000)]{2000A&AS..143...33B} Bonnarel, F., Fernique, P., Bienaym{\'e}, O., et al.\ 2000, \aaps, 143, 33 

\bibitem[Bosma(1981)]{1981AJ.....86.1825B} Bosma, A.\ 1981, \aj, 86, 1825 

\bibitem[Bottema(1993)]{1993A&A...275...16B} Bottema, R.\ 1993, \aap, 275, 16 

\bibitem[Bottema(1997)]{1997A&A...328..517B} Bottema, R.\ 1997, \aap, 328, 517 

\bibitem[Burkert(1995)]{1995ApJ...447L..25B} Burkert, A.\ 1995, \apjl, 447, 
L25 

\bibitem[Cenarro et al.(2001)]{2001MNRAS.326..959C} Cenarro, A.~J., 
Cardiel, N., Gorgas, J., et al.\ 2001, \mnras, 326, 959 

\bibitem[Cloet-Osselaer et al.(2012)]{2012MNRAS.423..735C} Cloet-Osselaer, 
A., De Rijcke, S., Schroyen, J., \& Dury, V.\ 2012, \mnras, 423, 735 

\bibitem[Courteau et al.(2014)]{2014RvMP...86...47C} Courteau, S., 
Cappellari, M., de Jong, R.~S., et al.\ 2014, Reviews of Modern Physics, 
86, 47 

\bibitem[de Souza et al.(2004)]{2004ApJS..153..411D} de Souza, R.~E., 
Gadotti, D.~A., \& dos Anjos, S.\ 2004, \apjs, 153, 411 

\bibitem[de Blok et al.(2008)]{2008AJ....136.2648D} de Blok, W.~J.~G., 
Walter, F., Brinks, E., et al.\ 2008, \aj, 136, 2648 

\bibitem[de Blok et al.(2001)]{2001AJ....122.2396D} de Blok, W.~J.~G., 
McGaugh, S.~S., \& Rubin, V.~C.\ 2001, \aj, 122, 2396 

\bibitem[De Geyter et 
al.(2013)]{2013A&A...550A..74D} De Geyter, G., Baes, M., Fritz, J., \& Camps, P.\ 2013, \aap, 550, A74 

\bibitem[de Jong 
\& Bell(2007)]{2007iuse.book..107D} de Jong, R.~S., \& Bell, E.~F.\ 2007, Island Universes - Structure and Evolution of Disk Galaxies, 107 

\bibitem[Di Cintio et al.(2014)]{2014MNRAS.437..415D} Di Cintio, A., Brook, 
C.~B., Macci{\`o}, A.~V., et al.\ 2014, \mnras, 437, 415 

\bibitem[D'Odorico et al.(1998)]{1998SPIE.3355..507D} D'Odorico, S., 
Beletic, J.~W., Amico, P., et al.\ 1998, \procspie, 3355, 507 

\bibitem[Donato et al.(2009)]{2009MNRAS.397.1169D} Donato, F., Gentile, G., 
Salucci, P., et al.\ 2009, \mnras, 397, 1169 

\bibitem[Fuchs(2003)]{2003Ap&SS.284..719F} Fuchs, B.\ 2003, \apss, 284, 719 

\bibitem[Gentile et al.(2004)]{2004MNRAS.351..903G} Gentile, G., Salucci, 
P., Klein, U., Vergani, D., \& Kalberla, P.\ 2004, \mnras, 351, 903 

\bibitem[Gentile et al.(2007)]{2007MNRAS.375..199G} Gentile, G., Salucci, 
P., Klein, U., \& Granato, G.~L.\ 2007, \mnras, 375, 199 

\bibitem[Gentile et al.(2009)]{2009Natur.461..627G} Gentile, G., Famaey, 
B., Zhao, H., \& Salucci, P.\ 2009, \nat, 461, 627 

\bibitem[Gentile et 
al.(2013)]{2013A&A...554A.125G} Gentile, G., J{\'o}zsa, G.~I.~G., Serra, P., et al.\ 2013, \aap, 554, A125 

\bibitem[Gerssen et al.(1997)]{1997MNRAS.288..618G} Gerssen, J., Kuijken, 
K., \& Merrifield, M.~R.\ 1997, \mnras, 288, 618 

\bibitem[Gerssen et al.(2000)]{2000MNRAS.317..545G} Gerssen, J., Kuijken, 
K., \& Merrifield, M.~R.\ 2000, \mnras, 317, 545 

\bibitem[Gerssen 
\& Shapiro Griffin(2012)]{2012MNRAS.423.2726G} Gerssen, J., \& Shapiro Griffin, K.\ 2012, \mnras, 423, 2726 

\bibitem[Governato et al.(2012)]{2012MNRAS.422.1231G} Governato, F., 
Zolotov, A., Pontzen, A., et al.\ 2012, \mnras, 422, 1231 

\bibitem[H{\"a}nninen 
\& Flynn(2002)]{2002MNRAS.337..731H} H{\"a}nninen, J., \& Flynn, C.\ 2002, \mnras, 337, 731 

\bibitem[Into 
\& Portinari(2013)]{2013MNRAS.430.2715I} Into, T., \& Portinari, L.\ 2013, \mnras, 430, 2715 

\bibitem[Jenkins 
\& Binney(1990)]{1990MNRAS.245..305J} Jenkins, A., \& Binney, J.\ 1990, \mnras, 245, 305 

\bibitem[J{\'o}zsa et 
al.(2007)]{2007A&A...468..731J} J{\'o}zsa, G.~I.~G., Kenn, F., Klein, U., \& Oosterloo, T.~A.\ 2007, \aap, 468, 731 

\bibitem[Kassin et al.(2006)]{2006ApJS..162...80K} Kassin, S.~A., de Jong, 
R.~S., \& Pogge, R.~W.\ 2006a, \apjs, 162, 80 

\bibitem[Kassin et al.(2006)]{2006ApJ...643..804K} Kassin, S.~A., de Jong, 
R.~S., \& Weiner, B.~J.\ 2006b, \apj, 643, 804 

\bibitem[Koleva et 
al.(2009)]{2009A&A...501.1269K} Koleva, M., Prugniel, P., Bouchard, A., \& Wu, Y.\ 2009, \aap, 501, 1269 

\bibitem[Kregel et al.(2002)]{2002MNRAS.334..646K} Kregel, M., van der 
Kruit, P.~C., \& de Grijs, R.\ 2002, \mnras, 334, 646 

\bibitem[Kregel et al.(2005)]{2005MNRAS.358..503K} Kregel, M., van der 
Kruit, P.~C., \& Freeman, K.~C.\ 2005, \mnras, 358, 503 

\bibitem[Kuzio de Naray 
\& Kaufmann(2011)]{2011MNRAS.414.3617K} Kuzio de Naray, R., \& Kaufmann, T.\ 2011, \mnras, 414, 3617 

\bibitem[Macci{\`o} et al.(2012)]{2012ApJ...744L...9M} Macci{\`o}, A.~V., 
Stinson, G., Brook, C.~B., et al.\ 2012, \apjl, 744, L9 

\bibitem[Martinsson et 
al.(2013)]{2013A&A...557A.131M} Martinsson, T.~P.~K., Verheijen,
M.~A.~W., Westfall, K.~B., et al.\ 2013, \aap, 557, A131 

\bibitem[Mathewson et al.(1992)]{1992ApJS...81..413M} Mathewson, D.~S., 
Ford, V.~L., \& Buchhorn, M.\ 1992, \apjs, 81, 413 

\bibitem[McGaugh 
\& Schombert(2014)]{2014arXiv1407.1839M} McGaugh, S., \& Schombert,
J.\ 2014, AJ in press (arXiv:1407.1839)

\bibitem[Minchev 
\& Quillen(2006)]{2006MNRAS.368..623M} Minchev, I., \& Quillen, A.~C.\ 2006, \mnras, 368, 623 

\bibitem[Moorwood et al.(1998)]{1998Msngr..91....9M} Moorwood, A., Cuby, 
J.-G., \& Lidman, C.\ 1998, The Messenger, 91, 9 

\bibitem[Neto et al.(2007)]{2007MNRAS.381.1450N} Neto, A.~F., Gao, L., 
Bett, P., et al.\ 2007, \mnras, 381, 1450 

\bibitem[Oh et al.(2011a)]{2011AJ....142...24O} Oh, S.-H., Brook, C., 
Governato, F., et al.\ 2011a, \aj, 142, 24 

\bibitem[Oh et al.(2011b)]{2011AJ....141..193O} Oh, S.-H., de Blok, 
W.~J.~G., Brinks, E., Walter, F., 
\& Kennicutt, R.~C., Jr.\ 2011b, \aj, 141, 193 

\bibitem[Palunas 
\& Williams(2000)]{2000AJ....120.2884P} Palunas, P., \& Williams,
T.~B.\ 2000, \aj, 120, 2884  

\bibitem[Pecaut 
\& Mamajek(2013)]{2013ApJS..208....9P} Pecaut, M.~J., \& Mamajek, E.~E.\ 2013, \apjs, 208, 9 

\bibitem[P{\'e}rez et 
al.(2004)]{2004A&A...424..799P} P{\'e}rez, I., Fux, R., \& Freeman, K.\ 2004, \aap, 424, 799 

\bibitem[Persic et al.(1996)]{1996MNRAS.281...27P} Persic, M., Salucci, P., 
\& Stel, F.\ 1996, \mnras, 281, 27 

\bibitem[Portinari et al.(2004)]{2004MNRAS.347..691P} Portinari, L., 
Sommer-Larsen, J., \& Tantalo, R.\ 2004, \mnras, 347, 691 

\bibitem[Salucci et al.(2007)]{2007MNRAS.378...41S} Salucci, P., Lapi, A., 
Tonini, C., et al.\ 2007, \mnras, 378, 41 

\bibitem[Sault et al.(1995)]{1995ASPC...77..433S} Sault, R.~J., Teuben, 
P.~J., 
\& Wright, M.~C.~H.\ 1995, Astronomical Data Analysis Software and Systems IV, 77, 433 

\bibitem[Sellwood 
\& Binney(2002)]{2002MNRAS.336..785S} Sellwood, J.~A., \& Binney, J.~J.\ 2002, \mnras, 336, 785 

\bibitem[Sellwood(2014)]{2014RvMP...86....1S} Sellwood, J.~A.\ 2014, 
Reviews of Modern Physics, 86, 1 

\bibitem[Shapiro et al.(2003)]{2003AJ....126.2707S} Shapiro, K.~L., 
Gerssen, J., \& van der Marel, R.~P.\ 2003, \aj, 126, 2707 

\bibitem[van Eymeren et 
al.(2009)]{2009A&A...505....1V} van Eymeren, J., Trachternach, C., Koribalski, B.~S., \& Dettmar, R.-J.\ 2009, \aap, 505, 1 

\bibitem[Vazdekis et al.(2010)]{2010MNRAS.404.1639V} Vazdekis, A., 
S{\'a}nchez-Bl{\'a}zquez, P., Falc{\'o}n-Barroso, J., et al.\ 2010, \mnras, 
404, 1639 

\bibitem[Weiner et al.(2001)]{2001ApJ...546..931W} Weiner, B.~J., Sellwood, 
J.~A., \& Williams, T.~B.\ 2001, \apj, 546, 931 

\bibitem[Zibetti et al.(2009)]{2009MNRAS.400.1181Z} Zibetti, S., Charlot, 
S., \& Rix, H.-W.\ 2009, \mnras, 400, 1181 









\end{thebibliography}
\end{document}